\documentclass[12pt,preprint]{aastex}
\usepackage{amsfonts}
\newcommand{\R}{\mathbb R}

\usepackage{graphics} 
\usepackage{rotating}
\begin{document}

\begin{center}
{\Large{Minimal Energy Transfer of Solid Material Between Planetary Systems}}\\
\medskip\medskip\medskip

Edward Belbruno$^a$, Amaya Moro-Mart\'{\i}n$^a$ and Renu Malhotra$^b$\\
\medskip\medskip

$^a$ Department of Astrophysical Sciences, Princeton University.\\
$^b$ Department of Planetary Sciences, University of Arizona.\\ 
\medskip
belbruno@princeton.edu, amaya@astro.princeton.edu, renu@lpl.arizona.edu\\

\end{center}
\medskip\medskip
\begin{center}
ABSTRACT
\end{center}

The exchange of meteorites among the terrestrial planets of our Solar System is a well established 
phenomenon that has triggered discussion of lithopanspermia within the Solar System. Similarly, could solid
 material be transferred across planetary systems? To address this question, we explore the dynamics of the 
transfer of small bodies between planetary systems. In particular, we examine a dynamical process that yields 
very low escape velocities using nearly parabolic trajectories, and the reverse process that allows for low 
velocity capture. These processes are chaotic and provide a mechanism for minimal energy transfer 
that yield an increased transfer probability compared to that of previously studied mechanisms that have 
invoked hyperbolic trajectories.  We estimate the transfer probability in a stellar cluster as a function of 
stellar mass and cluster size.  We find that significant amounts of solid material could potentially have been 
transferred from the early Solar System to our nearest neighbor stars.  While this low velocity mechanism improves 
the odds for interstellar lithopanspermia, the exchange of biologically active materials across stellar systems depends 
greatly upon the highly uncertain viability of organisms over the timescales for transfer, typically millions of years.

\newpage

\begin{center}
1. INTRODUCTION
\end{center}

\medskip

From the collection of thousands of meteorites found on Earth, there are about 20 that have been 
identified as having a Martian origin, and a similar number that originated from the 
Moon. The study of the dynamical evolution of these meteorites agrees well with the 
cosmic ray exposure time and with the frequency of landing. 
No meteorites were ever found on the Moon by the Apollo mission. In 2005, 
the rover Opportunity encountered in Mars the first meteorite on another Solar System body, 
identified as an iron-nickel meteorite. These findings, together with dynamical simulations
(Gladman 1997; Dones et al. 1999; Mileikowsky et al. 2000), indicate that
meteorites are exchanged among the terrestrial planets of our Solar System at a measurable level. 
Because sufficiently large rocks may protect dormant microorganisms from cosmic ray exposure
and from the hazards of the impact at landing, it has been suggested that 
the exchange of microorganisms living inside rocks could take place among the 
Solar System planets, a phenomenon known as {\it lithopanspermia}. 
In fact, new laboratory experiments have confirmed that several microorganisms (bacterial spores, cyanobacteria and lichen) 
embedded in martian-like rocks could survive under shock pressures similar to those suffered by martian meteorites upon impact ejection 
(St\"offler et al. 2007; Horneck et al. 2008). 
Under this scenario, life on Earth could potentially spread to other moons and planets within
our Solar System, and/or life on Earth could have an origin elsewhere
in our Solar System. 

\medskip

Melosh (2003) investigated the probability of lithopanspermia taking place
amongst the stars in the solar local neighborhood. He found that even though
numerical simulations show that up to one-third of all the 
meteorites originating from the terrestrial planets are ejected out of the Solar System 
by gravitational encounters with Jupiter and Saturn, the 
probability of landing on a terrestrial planet of a 
neighboring planetary system is extremely low because of the high relative velocities 
of the stars and the low stellar densities. He concluded that lithopanspermia 
among the current solar neighbors is ``overwhelmingly unlikely''.  

\medskip

In a subsequent paper, Adams \& Spergel (2005) pointed out that 
the majority of stars, including the Sun\footnote{In the Solar System, the existence of short-lived 
isotopes ($^{60}$Fe and $^{26}$Al) in primitive meteorites
has been interpreted as indication of nearby a supernova explosion shortly before the Solar System solids
started to accrete. This has been considered as evidence that the
Sun was born in a cluster environment.}, are born in stellar clusters with N = 100--1000 
members and a typical radius of $R = 1pc(N/100)^{1/2}$ (Lada $\&$ 
Lada 2003 and Carpenter 2000). In such an environment, the probability of transfer would be higher
due the larger stellar densities and smaller stellar relative velocities compared to those for field stars (and for the current solar neighborhood). 
The timescale for planet formation and the dispersal time of the clusters are comparable (10--100 Myr); 
therefore, it could be possible that solid material be transferred before the cluster disperses. 
Adam $\&$ Spergel (2005) estimated the probability of transfer of biologically-active
remnants between planetary systems within a cluster by using Monte Carlo simulations, 
assuming that the stars are mostly in binary systems (which increases the 
cross-section). Adopting typical ejection speeds of $\sim$5 km/s, they found that 
the expected number of successful lithopanspermia events per cluster is $\sim$ 10$^{-3}$; for 
lower ejection speeds, $\sim$ 2 km/s, this number is 1--2. 

\medskip

Because there is a significant increase in the number of possible lithopanspermia events with 
decreased ejection velocity, it is of interest to study a very low energy 
mechanism with velocities significantly smaller than those considered in Adams \& 
Spergel (2005). This mechanism was described by Belbruno (2004) in the mathematical
context of a class of nearly parabolic trajectories in the restricted three-body
problem. The escape velocities of these parabolic-type trajectories are very low, $\sim$ 0.1 km/s, 
substantially smaller than the mean relative velocity 
of stars in the cluster, and the remnant escapes the planetary system by slowly meandering away. 
This process of ``weak escape'' is chaotic in nature and its study requires the use of methods
of chaos theory, lying beyond the reach of Monte Carlo simulations
employed in previous studies. 

\medskip
``Weak escape" is a transitional motion between capture 
and escape. For it to occur, the trajectory of the remnant must pass near 
the largest planet in the system.  ``Weak capture'' is the reverse process, when a 
remnant can get captured with low velocity by another planetary system. The fact that the
escape velocities of the remnants we consider here are small, enhances the probability that
a remnant can be weakly captured by another planetary system due to 
lower approach velocities of the stars encountered.

\medskip

In $\S$ 2, $\S$ 3 and $\S$ 4, we describe the mathematics of weak escape and weak capture (based on Belbruno 2004), 
and in $\S$ 5 we disscuss its astronomical application to the study of the slow chaotic transfer of solid material (remnants) between planetary systems within a star cluster: $\S$ 5.1 describes the location of the region supporting weak capture and weak escape for a star of a given mass;  $\S$5.2 constraints the range of stellar masses that allow minimal energy transfer (weak transfer) to take place; $\S$5.3 estimates the probability of weak capture by a star of a given mass; and $\S$5.4 calculates the number of weak transfer events. The focus of the paper is the study of the transfer of solid material between planetary systems, without regard to whether or not they might contain material of biological interest. Our motivation, however, is to shed light on the following questions: could the building blocks of life on Earth have been transferred to other planetary systems within the first 10--100 Myr of the Solar System evolution, when the Sun was still embedded in its maternal aggregate?; and vice versa, could life on the Solar System have been originated beyond its boundaries? 
This is addressed in $\S$5.5, where we apply the results in the previous sections to lithopanspermia, 
focusing in particular on solar-type stars. 

\medskip

\begin{center}
2. MODEL
\end{center}

\medskip
We first describe the mathematics of weak escape and weak capture. 
Readers interested in the astronomical application only could proceed to $\S$5. 
We define a general planetary system, $S$, consisting of a central star, $P_1$, and a system of
$N$ planets, $P_i, i=2,...,N$ ($N \geq 3$) on co-planar orbits that are approximately 
circular. Their labeling is not reflective of their relative 
distances from $P_1$. We assume that the mass of the star, $m_1$, is much larger than the masses of 
any of the planets, $m_i$ ($m_1 \gg m_i, i=2,...,N$), and that the mass of one of
the planets, $P_2$, is much larger than the sum of the masses of all the other planets, 
$m_2 \gg ~m_i, i = 3,...,N$ (this condition is fulfilled in the case of our Solar System with $P_2=$~Jupiter). 
These assumptions reduce $S$ to a much simpler system without the loss of generality. 
We consider a remnant, $P_0$, whose mass, $m_0$, is negligible with respect to 
$P_i,~i=1,...,N$; we assume that $P_0$ orbits the star in the same plane as the planets without affecting
their orbits, and that it spends most of the time far from $P_1$.  Because $m_2 \gg m_i, i = 3,...,N$, 
the gravitational perturbation of $P_2$ on the motion of $P_0$ is the dominant one, and 
we will ignore the perturbations due to the other planets. This reduces the motion of $P_0$ 
to that of a three-body problem between $P_0, P_1$ and $P_2$, where $P_1$ and $P_2$ are moving in 
approximately circular orbits about their common center of mass. Because $m_1 \gg m_2$, 
we can view $P_1$ as fixed, with $P_2$ orbiting around it in a circular orbit at 
constant radial distance $\Delta$ ($1 AU \leq \Delta \leq 500 AU$) and orbital 
frequency $\omega$. This defines the classical ``planar circular restricted three-body problem'' 
for the motion of $P_0$.  

\medskip

The differential equations for the restricted three-body problem are well known. We write them in 
dimensionless form; the details can be found in Belbruno (2004). Without loss of generality, 
we choose a reference frame with origin at the center of mass of the $P_1$,$P_2$ system;
we choose units such that $\Delta = 1$, $\omega = 1$, $m_1 = 1-\mu$ and $m_2 = \mu$, where 
$\mu = m_2/(m_1+m_2) > 0$ and $\mu \ll 1$. 
Under these assumptions, the period of motion of $P_2$ around $P_1$ is $T=2\pi$. 
We will refer to this as the reduced Solar System, $S_0$. In our reduced Solar 
System, $P_1$ is the Sun, $P_2$ is Jupiter and $\mu \approx 0.001$. 
In inertial coordinates ($Q_1,Q_2$) the differential equations for the motion of $P_0$ can be written as
\begin{equation}
\ddot{Q} = {\Omega}_Q  ,
\label{eq:3}
\end{equation}
where 
we have used the notation $\dot{} \equiv \frac{d}{dt}$, and
$Q = (Q_1,Q_2)$, assumed to be a vector, and 
$\Omega_Q \equiv \frac{\partial\Omega}{\partial Q}$, with
\begin{eqnarray}
\Omega&=&\frac{1-\mu}{r_1}+\frac{\mu}{r_2},\\
r_1(t) &=& \sqrt{(Q_1 - \mu c)^2 +(Q_2 - \mu s)^2}, \\
r_2(t)&=&\sqrt{(Q_1 +(1- \mu) c)^2 +(Q_2 +(1- \mu) s)^2},
\end{eqnarray}
where $r_1$, $r_2$ are the distances of $P_0$ to $P_1$ and $P_2$, respectively, 
$c \equiv \cos(t), s\equiv \sin(t)$, and 
the position of $P_1$ is given by $\mu(c, s)$ and that of $P_2$ is given by
$-(1-\mu)(c,s)$.

\medskip

These differential equations can also be writen in a barycentric rotating coordinate system 
($x_1,x_2$) with orbital frequency $\omega$. In this case, they form an autonomous system with an energy integral, 
the Jacobi energy, $J = J(x,\dot{x})$, where $x=(x_1,x_2)$ is a vector and $J$ is a function on the 
four-dimensional phase space $(x,\dot{x})$. Along a solution, ($x(t),\dot{x}(t)$), $J$ is a 
constant of the motion and defines a three-dimensional surface constraining the motion of 
$P_0$ called the Jacobi surface. In the rotating system, $P_1$ and $P_2$ are fixed and it 
is convenient to place them on the $x_1$-axis, with $P_1$ at $x_1 = \mu$ and $P_2$ at 
$x_1 = -1+\mu$, in which case, 
\begin{equation}
J = -|\dot{x}|^2 + |x|^2 +\mu(1-\mu)+2\Omega .
\label{eq:2}
\end{equation}
where $|x|$ represents the standard Euclidian norm of $x$, and the additive 
term $\mu(1-\mu)$ is present so that the values of the Jacobi energy, $J = C$, are normalized.

\medskip

\begin{center}
3. WEAK CAPTURE AND ESCAPE
\end{center}

\medskip

In this section, we introduce the concepts of weak capture and escape (for a detailed discussion we 
refer to Belbruno (2004, 2007b). 
A convenient way to define the capture of $P_0$ with respect to $P_1$ or $P_2$ is by using the 
concept of ``ballistic capture''. 
We define the two-body Kepler energy, $E_k$, of $P_0$ with respect to one of the bodies $P_k, k=1,2$:
\begin{equation}
E_k = {1\over2}v^2 - {m_k\over r_k}
\end{equation}												
where $v$ is the velocity of $P_0$ relative to $P_k$.												
Ballistic capture takes place when $E_k \leq 0$.												
More precisely, let $\phi(t) = (Q(t), \dot{Q}(t))$ be a solution of Eq.(\ref{eq:3}) 							
for $P_0$, and assume no collisions take place, i.e. $r_k > 0$.  												
$P_0$ is ballistically captured by $P_k$ at time $t=t^*$ if $E_k(\phi(t^*)) \leq 0$; similarly,												
$P_0$ escapes $P_k$ at time $t=t_2$, if there exists a finite time interval $[t_1, t_2]$, with $t_2 > t_1$ 
where $E_k(\phi(t_1)) \leq 0$ and $E_k(\phi(t_2)) > 0$. 											
For notation, we set $E_k(\phi(t)) = E_k(t)$. 												

\medskip												

Now we consider the motion of $P_0$ around $P_2$ (this is known as Hill's problem). 												
If the motion of $P_0$ is initially circular and occurs sufficiently 												
close to $P_2$, it is found that the motion is generally stable and $E_2$ remains negative for 										 	 	
all times (this was proven by Kummer(1983) applying the Kolmogorov-Arnold-Moser theorem). 												
However, if $P_0$ is not sufficiently near to $P_2$, then the stability 												
will break down and the motion will substantially deviate from a circular orbit, 
leading to escape from $P_2$ due to the gravitational perturbation of $P_1$.  

Of particular interest is temporary ballistic capture. 
This occurs when $P_0$ is ballistically captured by $P_2$ over the finite interval $[t_1, t_2]$, i.e., 
when $E_2(t) \leq 0$, for $t_1 \leq t \leq t_2$, $E_2(t_1)=0, E_2(t_2) = 0$, and $E_2(t) > 0$ 
for $t < t_1$ and $t > t_2$. 
We say that $P_0$ is ``pseudo-ballistically captured'' at a time $t=t^*$ when $E_2(t^*) \stackrel{>}{\sim} 0$, i.e. 
where the Kepler energy is slightly hyperbolic. 

\medskip

Because temporary ballistic capture with respect to  $P_2$ is generally unstable and chaotic in nature 
(Belbruno 2004; Garcia \& Gomez 2007) and lies at the transition between capture and escape, it is referred to 
as ``weak capture''.  For weak capture to occur, the Jacobi energy $C$ must be small enough so that 
the velocity of $P_0$ is sufficiently large at a given distance from $P_2$ to be in this 
transition state. 

\medskip

As is evident by numerical integration of $P_0$ about $P_2$, 
weak capture generally occurs for relatively short time spans, $\Delta t=t_2-t_1$.
For example, in the case of the Earth-Moon system ($\mu = 0.012$), weak capture 
about the Moon occurs for time spans of days or weeks; in the case of the Sun-Jupiter system, 
the timescale is months or a few years (Belbruno \& Marsden 1997; Belbruno 2007a) 

\medskip

\begin{figure}[h!]
\begin{center}
\resizebox{90mm}{!}
{\includegraphics{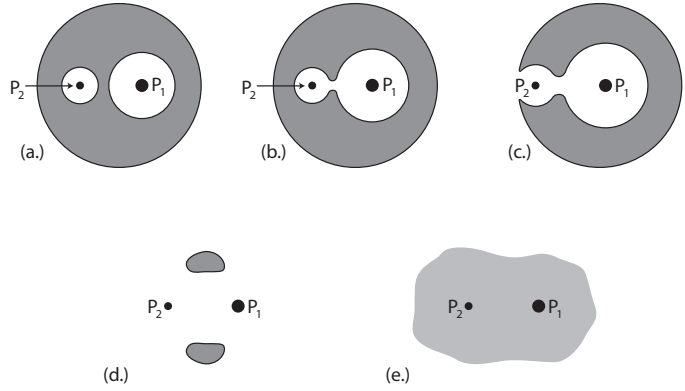}}
\end{center}

\caption{Basic Hill's regions (in {\it white}):
Starting from left to right and top to bottom, $C$ has the
values: $C > C_2; C_1 < C \stackrel{<}{\sim} C_2; C \stackrel{<}{\sim} C_1$; $C_3 < C < C_1$ and  $C < 3 $.}
\label{Figure:figureB}
\end{figure}

A numerical approach to estimate the region around the Moon supporting weak capture was 
developed by Belbruno (2004). This region was termed the weak stability boundary, $\mathcal{W}$, 
and it is the location in the phase space where the motion of $P_0$ with respect to $P_2$ lies between capture 
and escape. Recent results by Garcia \& Gomez (2007) have precisely determined  $\mathcal{W}$ using the numerical 
approach and have shown that
it is closely related to a complicated fractal region in phase space supporting unstable chaotic motion. However, 
the location of $\mathcal{W}$ can be can be approximately estimated analytically in a straight forward manner 
by a set $W$ described as follows:  

\noindent
Consider the barycentric rotating coordinates $x = (x_1,x_2)$. To be in 
weak capture, $P_0$ must have a suitably large velocity magnitude, $|\dot{x}|$, that depends on the 
velocity direction at a given distance $r_2$ from $P_2$. 
This is equivalent to the requirement that the Jacobi constant
$C$ needs to be in a suitably small range: we require that $C \stackrel{<}{\sim} C_1$, 
where $C_j$ is the value of $J$ at the classical Lagrange points $L_j$ (which are the locations 
at which $|\dot{x}| = 0$).  
We label the three collinear Lagrange points from left to right (in the rotating barycentric 
reference frame defined in $\S$ 2) with index 1,2, and 3; then the 
values of the Jacobi energy, $C_j$, at each of these points has the ordering: $C_2>C_1 =3$. 
As $C$ decreases, the allowed range of motion, $x$, of $P_0$ expands. 
When $C$ goes just below $C_1$, the Hill's region\footnote{The Hill's region for a given value of $C$ is 
the projection of the Jacobi surface onto the $x_1,x_2$ space, which yields locations where $P_0$ is allowed to move.} 
around $P_2$ opens near the position
of $L_1$ so that $P_0$ can move out of the Hill's region around $P_2$, and infinitely far away from 
both $P_1$ and $P_2$ 
(see Figure \ref{Figure:figureB} -- i.e. $C_1$ is the minimal Jacobi energy for escape). 
If we decrease $C$ further, $C<3 < C_1$, then the zero velocity curves disappear 
and the Hill's region becomes the entire $x_1x_2$-plane. 

\medskip

As is described in Belbruno (2004), the set $W$ has two components, $W_E$ and $W_H$, as follows.
\begin{eqnarray}
W &=& W_E \cup W_H, \\
W_E &=& \{ (x,\dot{x}) \in \R^4 |E_2 \leq 0, J = C, C^* \leq C < C_1, \dot{r}_2 = 0 \},\\
W_H&=& \{ (x,\dot{x}) \in \R^4 |E_2 \stackrel{>}{\sim} 0, J = C, C^* \leq C < C_1 \},
\end{eqnarray}
where the constant $C^*$ is determined so that $W$ exists and depends on the value of $\mu$. 
Weak capture occurs on $W_E$ (associated with osculating elliptic orbits about $P_2$), 
and pseudo-ballistic capture (equivalently pseudo-ballistic escape) occurs on 
$W_H$ (associated with slightly hyperbolic orbits), 

\medskip

For weak capture we need to consider the set $W_E$, 
\begin{equation}
W_E = \mathcal{J}(C) \cap \Sigma \cap \sigma,
\end{equation}
where $\mathcal{J} = \{ (x,\dot{x}) \in \R^4 | J = C \}$, 
$\Sigma = \{ (x,\dot{x}) \in \R^4 | E_2 \leq 0 \}$, $\sigma = \{ (x,\dot{x}) \in \R^4 | \dot{r}_2 = 0 \}$.    
$W_E$ is equivalent to a set of osculating elliptic orbits about $P_2$ with osculating pericenter 
$q_2$, in a two-dimensional annular region described by
\begin{equation}
q_2 = f(\theta_2, e_2),
\end{equation}
where $\theta_2$ ($0\leq\theta_2\leq2\pi$) is the polar angle measured with respect to the $P_1$,$P_2$ axis) 
in a $P_2$-centered rotating coordinate system, the function $f$ is periodic in $\theta_2$ of period $2\pi$, 
and $e_2$ is the osculating eccentricity of the orbit of $P_0$ with respect to  $P_2$ ($0 \leq e_2 \leq 1$).
For example, in the case where $C \stackrel{<}{\sim} C_1$ and $q_2 \stackrel{>}{\sim} 0$,$\mu \stackrel{>}{\sim} 0 $, 
 this relationship takes the form,
\begin{equation}
q_2 \approx
\frac{(1-e_2)\mu^{\frac{1}{3}}}
{3^{\frac{5}{3}} - \frac{2}{3} \mu^{\frac{1}{3}}}.
\end{equation}

\medskip

The set $W_H$ plays a key role on the existence of chaos, as it is proven that a hyperbolic invariant set 
associated to parabolic motion with respect to  $P_1$ exists on the set $W_H$, giving rise to chaotic 
dynamics (Belbruno 2004, pages 186-191).   

\medskip

A more accurate estimation of $\mathcal{W}$ has recently been obtained for a large range of the Jacobi energies 
by Belbruno et al. (2007b) through the visualization of special two-dimensional Poincar\'e sections, generally 
showing resonant tori lying within a chaotic sea. 
These sections yield a more accurate representation of $\mathcal{W}$, suggesting 
that it may consist of transverse homoclinic points, 
related to the intersections of the invariant manifolds of the Lyapunov periodic orbits associated to the 
unstable Lagrange points. 
It is demonstrated that during weak capture, $P_0$ moves around $P_1$ while 
transitioning between various resonant states with respect to  $P_2$. 
However, these results apply to elliptic-like resonant motions that are not sufficiently 
energetic for the purposes of this paper. In the following, we will be considering the parabolic motions 
associated with $W_H$.     

\medskip

In our Solar System, the existence of the weak stability boundary and the 
viability of weak capture was demonstrated in practice by the Japanese spacecraft {\it Hiten}: 
using a trajectory designed by Belbruno (1990, 1993, 2007a), {\it Hiten} was captured into an orbit 
around the Moon in 1991 without the use of rockets to slow down.  
Weak capture at the Moon was also achieved in 2004 by the ESA spacecraft {\it SMART1} 
(Racca 2003, Belbruno 2007a). In both cases, the capture was chaotic and unstable because 
the orbit lay in the transition between capture and escape. In another application, 
weak escape from the Earth's L$_4$~(or L$_5$) was invoked to suggest a low energy transfer to the 
Earth for the hypothetical Mars-sized impactor that is thought to have triggered the ``giant impact'' 
origin of the Moon (Belbruno \& Gott 2005). 

\medskip

\begin{center}
4. PARABOLIC MOTION AND CHAOS
\end{center}
\medskip

In this section, we describe the mechanism for low velocity escape from $S$ that is chaotic and applies to the set of 
parabolic trajectories around $P_1$. The reverse process yields low velocity chaotic capture into $S$.  
Consider a solution $Q(t)$ of (\ref{eq:3}) for $P_0$ in the case of the reduced Solar System $S_0$ for the 
restricted three-body problem: $P_0$ is on a parabolic trajectory with respect to  $P_1$ when $\lim_{t 
\rightarrow \pm \infty} |Q| = \infty$ and $\lim_{t \rightarrow \pm \infty} |\dot Q| = 0$. 
In the absence of any of the planets, the motion of $P_0$ is that of a standard parabola around $P_1$, and 
the Jacobi energy has the value $C = \pm 2\sqrt{2}$, where + is for retrograde parabolic trajectories, and -- is for 
direct trajectories. 

\medskip

In the case of the Solar System, $S$, because the masses of the planets are very small with respect to that of $P_1$, 
the motion of $P_0$ will be slightly perturbed and the trajectories that lie very close to parabolic trajectories will 
be able to escape $P_1$ with very small velocities; we refer to these as ``pseudo-parabolic trajectories''.  
These trajectories satisfy $ |Q|  \rightarrow R, \ \  |\dot Q| \rightarrow \sigma,$ as $t \rightarrow T_R$, and, 
without loss of generality, we assume that $P_0$ started at periapsis with respect to  $P_1$ at time $t=0$ and traveled a 
distance $R$ in a sufficiently long time $T_R > 0$.
The value of $\sigma$ is small and coincides with the escape velocity with respect to  $P_1$ at the distance $R$, 
$\sigma = \sqrt{2Gm_1/R}$; $R$ is chosen sufficiently large so that $P_0$ can escape $S$ when it has a velocity 
close to $\sigma$.

\medskip

For notational purposes, the term ``parabolic trajectory'' refers to the solution of (\ref{eq:3}) for $S_0$ in 
scaled dimensionless coordinates and the precise definition of parabolic motion; whereas, 
when using the term ``pseudo-parabolic trajectory'', we are considering system $S$ in unscaled dimensional 
coordinates.  We now describe the main results for $S_0$, followed by the results for $S$. 
\medskip

\noindent
{\it Main Results for $S_0$}: 

\medskip 

Let $\phi(t) = (Q(t), \dot{Q}(t))$ be a parabolic solution of (\ref{eq:3}) for $P_0$ in a system where $\mu = 0$ 
and where $P_0$ does not collide with $P_1$ (i.e.~$r_1 > 0$). Consider the set of parabolic trajectories 
with $|C| \stackrel{<}{\sim} 2\sqrt{2} \approx 2.83$, that pass through the location of 
$P_2$ (now with zero mass because $\mu = 0$). Belbruno (2004) found that for $0 \leq \mu \ll 1$, 
if $P_0$ passes sufficiently close to $P_2$ without collision ($r_2 > 0$) and slightly 
beyond the distance $\Delta$, then: (a) when $P_0$ passes close to $P_2$, it is slightly 
hyperbolic and in the set $W_H$ (i.e. pseudo-weakly captured) around $P_2$; (b) $P_0$ never 
collides with $P_1$ or $P_2$; (c) the periapsis location of $P_0$ with respect to  $P_2$ is 
approximately the periapsis location with respect to  $P_1$; (d) there exists a set of positive 
measure of parabolic trajectories (i.e. parabolic trajectories exist); and (e) the motion of $P_0$ is chaotic. 

There are two possible outcomes: 
\begin{enumerate}
\item $P_0$ parabolically escapes $P_1$, i.e. $|Q| \rightarrow \infty$ and $|\dot{Q}| \rightarrow 0$ 
as $t \rightarrow \infty$. 
\item $P_0$ does not escape but instead, after traveling a large finite distance, 
falls back towards $P_1$, passing again by $P_2$ in a slightly hyperbolic orbit w.r.t $P_2$ in 
which $P_0$ is pseudo-weakly captured by $P_2$, after which it flies around $P_1$ and out again toward $|Q|=\infty$. 
\end{enumerate}

In the second case, every time $P_0$ falls back to fly by $P_2$, the argument of periapsis of the 
pseudo-parabola with respect to  $P_1$ is chaotic in nature and can take on a random value. 
If $P_0$ were to keep falling back to $P_2$ again and again, it would be permanently captured by $P_1$. 
However, it can be proven that this does not happen (in mathematical terms this is because the set of orbits 
leading to permanent capture are of zero measure); on the contrary, escape eventually takes 
place (i.e. there is a set of positive measure of parabolic trajectories that will escape to 
infinity after flying by $P_2$). 
In addition, in this case, there is a constraint on the fly-by distance $q_2$ as a function of $\mu$. 
If we consider the case of a direct fly by, the velocity of $P_0$ with respect to  $P_1$ when it 
flies by $P_2$ is $V$ $\approx$ 1.414 (in dimensionless units) -- this is because the fly-by 
is close to $P_2$ and therefore has a distance $\Delta \approx 1$ to $P_1$, and 
hence, $V \approx \sqrt{2/\Delta} \approx 1.414$); therefore, the velocity of $P_0$ with 
respect to  $P_2$ is approximately 0.414, and the Kepler energy of $P_0$ with 
respect to  $P_2$ is $E_2 \approx 0.08 - (\mu/r_2)$. 
A slight hyperbolic fly-by implies that $E_2 \stackrel{>}{\sim} 0$, or 
equivalently, $r_2 \stackrel{>}{\sim} {\mu \over 0.08}$. 
In the case of the Solar System, if $P_2$ is 
Jupiter ($\mu$ = 0.001) then $r_2 \stackrel{>}{\sim}$ 0.013 = 10,172,800 km, 
while for Neptune ($\mu$ = 0.00005), $r_2 \stackrel{>}{\sim}$ 0.0006 = 2,692,800 km.

\medskip
\noindent {\it Main Result for $S$}:
\medskip

The parabolic motion previously described for $P_0$ in the system $S_0$ can be translated to the pseudo-parabolic 
motion of $P_0$ in the system $S$ in the following manner. 
There is a set of positive measure of pseudo-parabolic trajectories with respect to  $P_1$ that 
pass by $P_2$ without colliding with it, and where $P_0$ is slightly hyperbolic with respect to $P_2$. 
The motion of such trajectories is sensitive due to the fly-by. As $P_0$ approaches $P_2$, it achieves a 
periapsis with respect to $P_2$ slightly beyond $P_2$'s orbit, with energy at periapsis slightly hyperbolic with 
respect to  $P_2$.  At that location, it is also approximately at the periapsis with respect to  $P_1$. $P_0$ then 
flies outward away from $P_1$ for a long period of time $T_R$, reaching a distance $R$ with a velocity $\sim\sigma$. 
$P_0$ pseudo-parabolically escapes $P_1$ with approximate parabolic escape velocity magnitude $\sigma$, when $r_1$ 
increases beyond $R$ as $t$ increases beyond $T_R$. 

\medskip

\noindent {\it Weak Stability Boundary Around $P_1$}  
\medskip

The pseudo-parabolic escape of $P_0$ occurs for $R$ sufficiently large so that the gravitational 
perturbation in $P_2$ is negligible. We have a two-body problem between $P_0$ and $P_1$, where $P_0$ 
is moving with a small velocity $\sigma$. We can define a weak stability boundary about $P_1$. This can 
be accomplished by considering a situation 
where the star $P_1$ is not isolated but included within a star cluster. We can think of a new ``three-body problem'' 
consisting of $P_1$, $P_0$, and a third object, $B$, that mimics the combined gravitational forces from the neighboring 
stars in the cluster. The gravitational perturbation of $B$ together with the gravity of $P_1$ forms a weak stability 
boundary region far from $P_1$ that is located at a finite distance that depends on the characteristics of the 
cluster and the separation of the stars. $P_0$ lies in the transition between capture and escape from $P_1$, 
when $R$ is sufficiently large and $\sigma$ is sufficiently small.

\begin{center}
5. REMNANT TRANSFER BETWEEN STELLAR SYSTEMS IN OPEN STAR
CLUSTERS
\end{center}
\medskip

Using the framework discussed in the previous sections we now consider the problem of remnant transfer between 
planetary systems in star clusters with low velocity dispersion, 
since low relative velocities are required for the weak capture mechanism. 
Specifically, we consider open clusters which typically have low relative stellar velocities, 
$U$ $\approx$ 1 km/s. For comparison, we note that in older globular 
clusters, stellar velocity dispersion can reach many 10's of km/s.  
Also for comparison, we mention that when a cluster starts to disperse, 
the relative distances and relatives velocities between the Sun and neighboring stars increase.  
For example, in the solar neighborhood, the Sun's closest neighbor $\alpha$-Centauri is $2.6\times10^5$ AU away 
(1.28 pc) with a relative velocity of 6 km/s. The latter is significantly higher than the $\sim 1$~km/s 
required for weak capture, making the transfer of material between the two stars very unlikely via weak transfer. 
Therefore, we will consider relatively young open clusters.
\medskip

We assume that the stars in the open cluster are approximately uniformly spaced in a three-dimensional grid 
by a distance $D$ and that their distribution is isotropic. Imagine a remnant ($P_0$) in a 
planetary system ($S$) passing near the primary planet ($P_2$) having a weakly hyperbolic flyby, moving away 
from the star ($P_1$) on a pseudo-parabolic trajectory, and reaching a distance $R = R_{esc}(m_1)$ from the star. 
At this distance, the small gravitational force acting on the remnant due to the central star is roughly 
comparable with the resultant gravitational force from the other stars in the cluster, as described above; as a consequence, 
the motion of the remnant becomes unstable, with small changes to its velocity 
leading to large changes in its trajectory that can either lead to capture or escape from the 
central star\footnote{The sensitivity of the motion of $P_0$ at the distance $R$ can be deduced from an 
analogous four-body problem described in Belbruno (2004) and Marsden \& Ross (2006) for a transfer to the Moon used 
by the spacecraft {\it Hiten}. 
In this case, the four bodies are the Earth ($P_1$), the Moon ($P_2$), the Sun ($P_3$) and the spacecraft ($P_0$).  
The spacecraft leaves the Earth and travels out to roughly 1.5$\times$10$^6$ km where the gravitational force of the 
Sun acting on the spacecraft approximately balances that of the Earth. At this location, the motion of the spacecraft 
is highly sensitive to small differences in velocity, and lies between capture and escape from the Earth, i.e. 
lies at the weak stability boundary between the Earth and Sun. The spacecraft then falls back towards the Earth 
and the Moon entering the weak stability boundary between the Moon and the Earth, which finally leads to capture by 
the Moon.}. 
In other words, the sphere of radius $R$ around the central star lies within a weak stability region and can 
be thought of as a uniform slice through the more complicated weak stability region. 

\medskip

\begin{figure}[h!]
\begin{center}
\resizebox{60mm}{!}
{\includegraphics{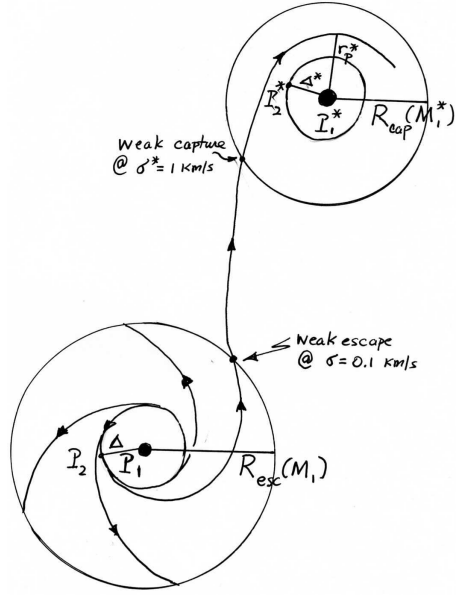}}
\end{center}
\caption{Trajectory weakly escaping $P_1$ at a distance $R=R_{esc}(m_1)$, being 
weakly captured by $P_1^*$ at a distance $R = R_{cap}(m_1^*)$, 
and moving to periapsis with respect to $P_1^*$ (motion projected onto a plane).}
\label{Figure:Transfernew}
\end{figure}

As shown in Figure \ref{Figure:Transfernew}, due to the structure of the space of parabolic trajectories, 
for any point $p$ on the circle of radius $R_{esc}(m_1)$ around the star $P_1$, there is a 
weakly-escaping trajectory that will pass by $p$ with velocity $\sigma$ moving away from the 
star (see also Figure 1 in Moro-Mart\'{\i}n $\&$ Malhotra 2005). Assume that the 
point $p$ lies on the line between the two stars. As the remnant ($P_0$) passes 
through this point and moves beyond the distance $R_{esc}(m_1)$ from the star, 
it goes beyond its weak stability boundary, moving to a region where the gravitational force of 
the star is negligible. The trajectory then continues undisturbed until the remnant moves within 
the weak stability boundary of another star, $P_1^*$ in planetary system $S^*$, 
located at a distance $R_{cap}(m_1^*)$ from $P_1^*$.  The remnant can then get 
captured by $P_1^*$, with a periapsis distance that is minimally a collision or maximally a 
distance $R_{cap}(m_1^*)$ from $P_1^*$. 
\medskip

Analogous to planetary system $S$, we are assuming that $S^*$ has a dominant planet $P_2^*$ at a 
radial distance $\Delta^*$ from the star $P_1^*$ (see Figure \ref{Figure:Transfernew}). 
For weak escape to occur, it is necessary that the remnant $P_0$ weakly escapes the star $P_1$ in the same 
plane of motion as that of the dominant planet $P_2$, and that the periapsis distance, $r_p$, of $P_0$ with 
respect to $P_1$ is approximately the semi-major axis of the dominant planet, $\Delta \approx r_p$. 
However, when $P_0$ is weakly captured by $P_1^*$ and moves to periapsis distance $r_p^*$ with 
respect to $P_1^*$ ($ 0 \leq r_p^* \leq R_{cap}(m_1^*)$), it need not approach this periapsis 
within the same plane of motion as the dominant planet $P_2^*$. This capture into $S^*$ is 
three-dimensional in nature and the remnant $P_0$ can approach the star $P_1^*$ from any direction. 
\medskip

The construction of the transfer from the periapsis with respect to $P_1$ (and also $P_2$)
to the periapsis with respect to $P_1^*$ can be approximated by determining a solution for $P_0$ in a five-body problem
between $P_0, P_1, P_2, P_1^*$ and $P_2^*$. We can approximate it by 
piecing together the solutions between two three-body problems, $P_0,P_1, P_2$ (hereafter TB1), 
and $P_0,P_1^*,P_2^*$ (hereafter TB2). This is done as follows. 
We know that pseudo-parabolic trajectories in the first three-body problem, $TB1$, exist as described above. 
In fact, the theory of their existence proves 
there is an infinite number of them, forming the structure of a Cantor set, that will travel out to a 
distance $R_{ecs}(m_1)$ where the gravitational perturbation due to $TB2$ begins to be felt and where 
the dynamics of $P_0$ becomes sensitive. The velocity of $P_0$ at this distance is approximated by the 
escape velocity from
$P_1$. That is, $P_0$ weakly escapes the first system. Let's assume that $P_0$ is then weakly captured by $TB2$. Its
velocity magnitude would then approximately  be the escape velocity from $P_1^*$ at the given distance, $R_{cap}(m_1^*)$.  
If $P_0$ were moving in the same plane as $P_2^*$ about $P_1^*$, then 
by the same theory on pseudo-parabolic trajectories, there would exist, by symmetry, an infinite number of trajectories 
that are weakly captured 
into the system $TB2$ and fly by $P_2^*$ in slight hyperbolic state. Then, a capture trajectory can be 
generated in forwards time as an extension of the the weak escape trajectory starting at a distance 
$R_{cap}(m_1^*)$ from $P_1^*$. 
If, on the other hand, the trajectory of $P_0$ does not lie in the same plane of motion of $P_2^*$ when it is weakly 
captured by $P_1^*$, then the theoretical results on pseudo-parabolic trajectories cannot be applied. 
However, the trajectory 
of $P_0$ can still be extended in forwards time and will move towards $P_1^*$ and fly by $P_2^*$. 
In this case, it cannot be guaranteed that $P_0$ will fly by $P_2^*$ in a weakly hyperbolic manner, 
although it would be reasonable that this would occur if the time of capture into $TP2$ were sufficiently long. 
More precisely, one of the following would occur:
\medskip

\noindent (i) $P_0$ remains in a bounded region about $P_1^*$ without 
collision with $P_1^*$. Then by the Poincar\'e recurrance theorem, $P_0$ would eventually fly by $P_2^*$ in a 
slight hyperbolic fashion or collide with $P_2^*$ 

\noindent (ii) $P_0$ collides with $P_1^*$ 

\noindent (iii) $P_0$ is ejected from the 
$P_1^* - P_2^*$- system.
\medskip

The piecing together of solutions at the weak stability 
boundary regions, for different types of trajectories, is discussed in Belbruno 2004 and Marsden $\&$ Ross 2006). 									 
The piecing together occurs at the weak stability boundaries of  $TB1$ and  $TB2$. 
\medskip									

\begin{center}									
5.1 Location of the Weak Stability Boundary									
\end{center}									
\medskip									

We now calculate $R_{esc}(m_1)$ and $R_{cap}(m_1^*)$ as a function of the stellar mass. 									
For weak escape to take place, the velocity, $\sigma$, of the remnant at the distance 									
$R_{esc}(m_1)$ from the star $P_1$ must be sufficiently smaller than the escape velocity at that distance 
from $P_1$ as well as the other stars in the cluster; i.e. it is on the weak stability boundary. 
At this distance the gravitational forces from $P_1$ and from the other stars in the cluster are comparable, 
and the motion of the remnant is unstable and chaotic in nature. 									
Because we are considering slow transfer within an open cluster with a characteristic dispersion 
velocity $U \approx 1$ km/s, we require that $\sigma$ is significantly smaller than $U$, 
i.e. of the order of 0.1 km/s. This is much smaller 									
than the nominal values of several km/s used by the Monte Carlo methods in 
previous studies (e.g.~Melosh 2003, Adams \& Spergel 2005). 									
\medskip

\begin{figure}[h!]
\begin{center}
\resizebox{60mm}{!}
{\includegraphics{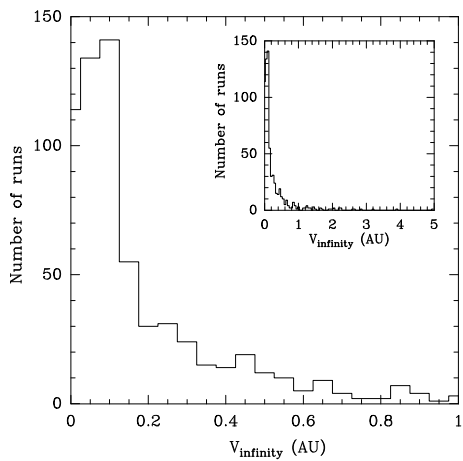}}
\end{center}
\caption{Velocity distribution of weakly escaping test particles from the Solar System.  
The model is a planar circular restricted three-body problem between the Sun, Jupiter and a massless particle.} 
\label{Figure:vhist}
\end{figure}

To place the above choice for $\sigma$ in context, we study the velocity 
distribution of weakly escaping test particles from the Solar System, using a three-body problem 
between the Sun ($P_1$), Jupiter ($P_2$) and a massless particle ($P_0$). To be consistent with our framework, 
we model this as a planar circular restricted three-body problem, where the test particle moves in the same plane 
of motion as Jupiter, assumed to be in a circular orbit at 5 AU. The trajectory 
of the test particle is numerically integrated by a standard Runge-Kutta scheme of 
order six and numerical accuracy of $10^{-8}$ in the scaled coordinates.  
The initial conditions of the test particle is an elliptic trajectory very close to parabolic with 
periapsis distance $r_p$=5 AU and apoapsis distance $r_a$=40,000 AU.  (Note that such orbits are not dissimilar 
to those of known long period comets in the Solar System.) 
For each numerical integration, we assume that Jupiter is at a random point in its orbit when the 
test particle starts from apoapsis 
at 40,000 AU and falls towards $P_1$. We record the time at which the test particle achieves escape with respect to the Sun 
(i.e. when the Kepler energy with respect to the Sun is positive) after performing a sufficient number of Jupiter fly-bys, 
and we record the resulting hyperbolic 
excess velocity $v_\infty$. 
Figure \ref{Figure:vhist} shows the distribution of the $v_\infty$: out of 670 cases, 
58\% have $v_\infty \leq$ 0.1 km/s and 79\% have $v_\infty \leq$ 0.3 km/s. 
Based on these results we will assume the velocity $\sigma$ of the remnant at the distance $R_{esc}(m_1)$ 
from the star to be in the range 0.1--0.3 km/s. 
\medskip

\begin{figure}[h!]
\begin{center}
\resizebox{120mm}{!}
{\includegraphics{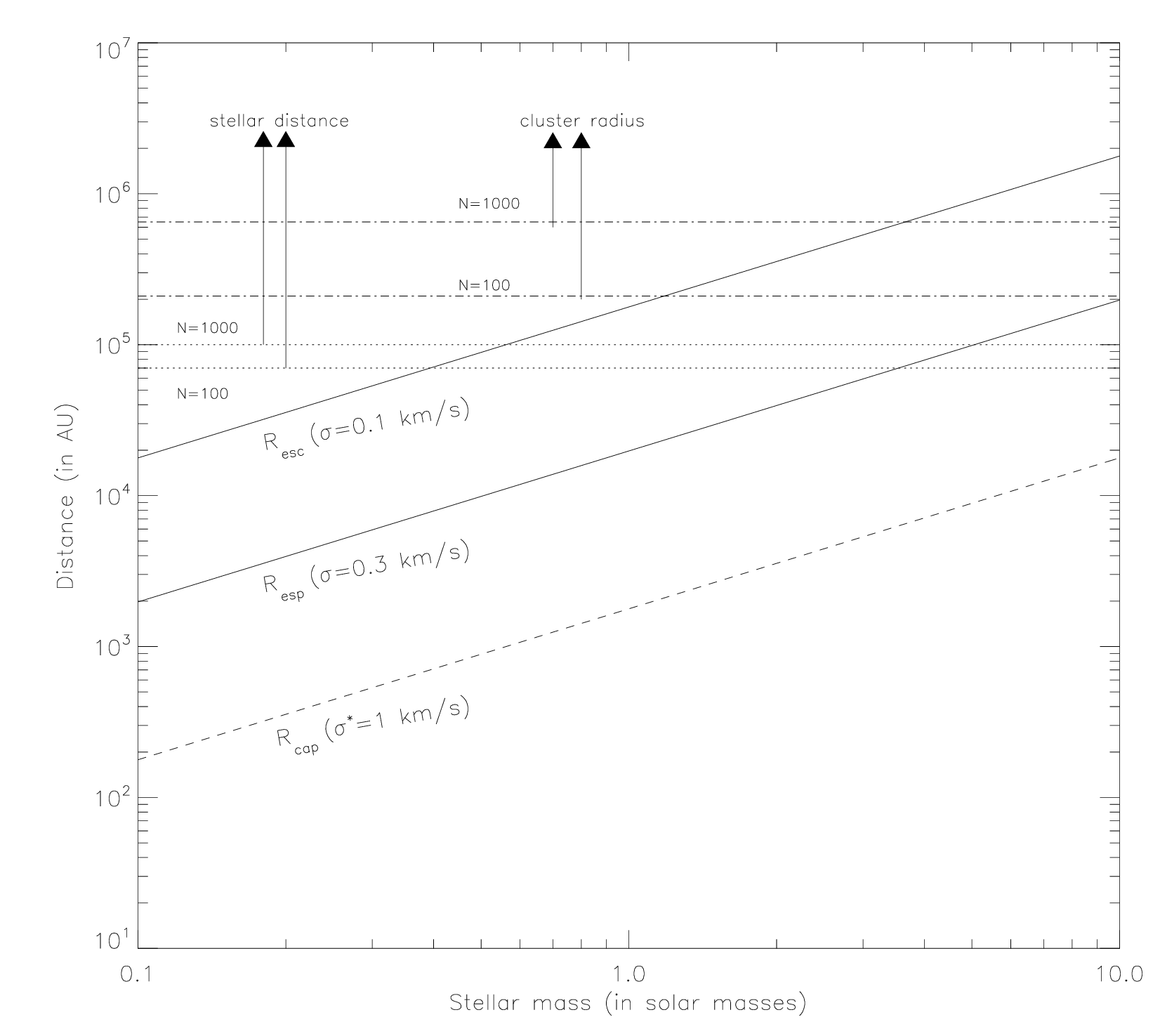}}
\end{center}
\caption{Radius of the weak stability boundary as a function of stellar mass. 
The weak escape boundary is defined as $R_{esc}(m_1) = 2Gm_1/\sigma^2$ with $\sigma = 0.1$--$0.3$~km/s, 
and the weak capture boundary is at $R_{cap}(m_1^*) = 2Gm_1^*/U^2$ with $U = 1$~km/s. 
The horizontal dashed-dotted line indicates the range of cluster sizes, while the dotted line indicates the the mean interstellar distance
in clusters consisting of N=100 and N=1000 members.}
\label{Figure:escape}
\end{figure}

For a given $\sigma$, the location of the weak stability boundary is approximately given 
by $R_{esc}(m_1) = 2Gm_1/\sigma^2$, where $m_1$ is the mass of the star and $\sigma$ is in the 
range 0.1--0.3~km/s (see Figure \ref{Figure:escape}). Beyond this boundary, we assume that the remnant 
will move at a constant velocity $\sigma$ with respect to the star.
\medskip

To allow slow chaotic transfer to a neighboring planetary system, the remnant needs to arrive at 
the distance $R_{cap}(m_1^*)$ with a relative velocity with respect to the target star ($P_1^*$) that 
is similar or smaller than its parabolic escape velocity at that distance, $\sigma^* = \sqrt{2Gm^*_1/R_{cap}(m_1^*)}$, 
where $m^*_1$ is the mass of the target star;  
if its velocity is higher than $\sigma^*$ it will not be captured and and will fly by.  
Since the relative velocity between stars in the cluster is $U \approx 1$~km/s, 
the remnant that weakly escaped from star $P_1$ moves toward the target star $P_1^*$ with 
velocity $U \pm \sigma$ (see Figure 
\ref{Figure:Transfernew}). Because $\sigma$ is small relative to $U$ and 
can be neglected, the relative velocity of the remnant with respect to the target
star is $\approx U$. Therefore, weak capture can occur at the distance at which 
$\sigma^* \approx U \approx 1$~km/s, i.e. $R_{cap}(m_1^*) = 2Gm_1^*/U^2$. 
\medskip

Figure \ref{Figure:escape} shows $R_{esc}(m_1)$ and $R_{cap}(m_1^*)$ as a function of the 
stellar mass. The horizontal dotted lines indicate the range of cluster sizes and mean 
interstellar distances for clusters consisting of 100 and 1000 members, respectively; 
such clusters are the birthplaces of a large fraction of stars in the Galaxy. The radius of the cluster 
depends on the number of stars, $N$, and is given by
\begin{equation}
R_{\rm cluster} = 1pc (N/100)^{1/2}.
\end{equation}
$R_{\rm cluster}$ is about $2.1\times10^5$ and $6.5\times10^5$ AU for 
N=100 and N=1000 members, respectively
(based on data in Lada $\&$ Lada 2003 and Carpenter 2000, and following Adams \& Spergel 2005). 
We can estimate the average interstellar distance within a cluster, 
\begin{equation}
D = n^{-1/3},
\end{equation}
where
$n = 3N/(4\pi R_{\rm cluster}^3)$ is the average number density of stars in the cluster.  
$D$ is about 7$\times$10$^4$ AU and 10$^5$ AU for a cluster with N=100 and N=1000 members, respectively.  
\medskip

\begin{center}
5.2 Constraints on Stellar Masses for Weak Transfer
\end{center}
\medskip

From Figure \ref{Figure:escape} we can set constraints on the stellar mass $m_1$ that could allow weak 
escape from $P_1$ to take place. The idea is simple: if for a given $\sigma$ (which as we saw in $\S$ 5.1 is in 
the range 0.1--0.3 km/s), $R_{esc}(m_1)<D$, i.e.~the weak stability boundary is located within the distance of 
the next neighboring star $P_1^*$, then weak transfer is possible because at the time the remnant passes near the 
star $P_1^*$, its velocity is similar to the mean stellar velocity dispersion, $U$ (which is a low $\sim1$~km/s in 
open clusters), and there is a significant probability of capture (which we quantify in the next section). 
Conversely, neighbor transfer by the process schematically represented in Figure \ref{Figure:Transfernew} is much 
less likely to take place if $R_{esc}(m_1)> D$ because at the time the remnant passes near a neighboring star, 
its velocity is too high and as a consequence it will simply fly by.

Consider $\sigma=0.1$~km/s.  Figure \ref{Figure:escape} shows that for clusters with 100 members, 
the condition $R_{esc}(m_1)< D$ for weak escape is satisfied for $m_1<0.4M\odot$, and for clusters 
with 1000 members, the condition is satisfied $m_1<0.55M\odot$.  
If we consider the higher but still acceptable value $\sigma=0.3$~km/s, Figure \ref{Figure:escape} 
shows that the stellar mass limits for weak escape are $m_1<3.5M_{\odot}$ and $m_1<5M_{\odot}$ for 
clusters of 100 members and of 1000 members, respectively. 
A remnant escaping parabolically from any star with larger mass than these limits will achieve a 
velocity of $\sigma<0.1$--0.3~km/s only at a distance larger than the mean interstellar distance 
in the cluster and weak transfer is not likely under such conditions.

\medskip

Of particular interest is the case of the Sun as the source of the remnants. It has been estimated 
that the Sun's birth cluster consisted of $N=2000\pm1100$ members (Adams \& Laughlin 2001).  
For a $1 M_{\odot}$ star in such a cluster, we find that the parabolic escape velocity at the 
mean interstellar distance is $\sigma\simeq0.12$--0.14 km/s.  These values of $\sigma$ 
certainly lie within the range of values of interest for weak escape. We conclude 
that {\it remnants originating in the early Solar System could in principle have met the 
conditions for weak escape in the Sun's birth cluster.} 
\medskip

\begin{center}
5.3 Probability of Weak Capture
\end{center}
\medskip

We have established in $\S$5.2 the range of stellar masses that could in principle allow weak escape to 
take place, we now estimate {\it roughly} the probability of capture by a neighboring planetary system. 

As mentioned previously, because the trajectory of the remnant approaching the target system $S^*$ is 
not necessarily in the same plane as that of the orbit of its primary planet $P_2^*$, and because the 
capture can be complicated (where $P_0$ could be captured for millions of years, moving in a complicated 
trajectory without planetary collisions), capture is not guaranteed even if the renmant $P_0$ falls within 
the weak stability boundary of the target star $P_1^*$. 
However, because a necessary condition for capture is that the renmant lies within this weak stability boundary, 
we can use the following geometrical considerations to set an {\it upper limit} for the capture probability. 
Whether or not the transfer of remnants from one star to any star of mass $m_1^*$ takes place will depend on: 
\begin{enumerate}

\item {\it The relative capture cross-section of the target star}. 

This is given by $ C_S = {G_f}(R_{cap}(m_1^*)/D)^2$, 
where $D$ is the distance between the two stars and $R_{cap}$ is 
described in $\S$5.1 and Figure \ref{Figure:escape} (dashed line). The factor $G_f$ represents the gravitational focusing, given by
\begin{equation}
G_f = 1 + ({v_{esc} \over v_{\infty}})^2,
\end{equation}
where $v_{\infty}$ is the velocity at infinity and $v_{esc}$ is the escape velocity at the distance $R_{cap}(m_1^*)$ from $P_1^*$.

In the case of weak capture, the term $G_f$ increases the cross-section due to enhanced gravity focusing.  For example, 
as we saw previously in the case of the Sun and Jupiter, 
$v_{\infty} \approx 0.1-0.3$ km/s, while at the distance $R_{cap} = 40,000$ AU, $v_{esc} \approx 0.2$ km/s. Therefore, in this example, 
$G_f \approx 2$, which doubles the capture cross-section. 
This situation occurs in our capture methodology. In the formulation for determining $R_{cap}(m_1^*)$, the remnant has an approximate 
relative 
approach velocity to the target star $P_1^*$ of roughly $U = 1$ km/s, which represents the $v_{\infty}$. However, $R_{cap}$ is determined 
so that this same value of velocity is taken for $v_{esc}$ from the target star. That is, we are assuming, 
$v_{\infty} \approx v_{esc} \approx 1$ km/s. This implies also that $G_f = 2$. This is a conservative estimate 
and does not make use of the nature of weak capture dynamics. In this situation, we have a trajectory with a 
$v_{inf} = 1$ with respect to $P_1^*$, which goes to a parabolic state with respect to $P_1^*$. 

As noted, the value $G_f = 2$ is conservative. In principle, the value of $G_f$ could be substantially increased 
if at a given value of $R_{cap}(m_1^*)$, $v_{\infty}$ is  smaller than the approximate value of $U = 1$ km/s, while 
the value of $v_{esc}$ remains the same as $U$. Dynamically, the way to decrease the $v_{\infty}$ with respect to 
$P_1^*$, as the remnant $P_0$ approaches $P_1^*$, is for $P_0$ to decrease its relative velocity. This process 
has been shown to exist in other problems; for example, in the case where we have a spacecraft, $P_0$, 
transferring from the Earth to the Moon on a trajectory that goes to ballistic capture at a given distance 
from the Moon (demonstrated by spacecraft as was discussed previously). Figure \ref{Figure:WSBTransfer}, 
shows a trajectory to the Moon, going to a periapsis distance of $500$ km after 111 days. When it arrives 
it’s velocity is approximately $v_{esc}$. However, it’s $v_{\infty}$ goes from a value of $1$ km/s to $0$. 
This is reflected in Figure \ref{Figure:VinfPlot} of the Kepler energy $K_M$ of $P_0$ with respect to the Moon 
along this transfer. At a sufficiently far distance from the Moon, where $v_{\infty} \approx \sqrt{2K_E}$, 
$v_{\infty}$ approaches zero. In the future, we would like to examine the distribution of the $v_{\infty}$ 
in the WSB about $P_1^*$ and thereby get a better understanding of the $G_f$ and the enhanced capture probabilities.

\begin{figure}[h!]
\begin{center}
\resizebox{100mm}{!}
{\includegraphics{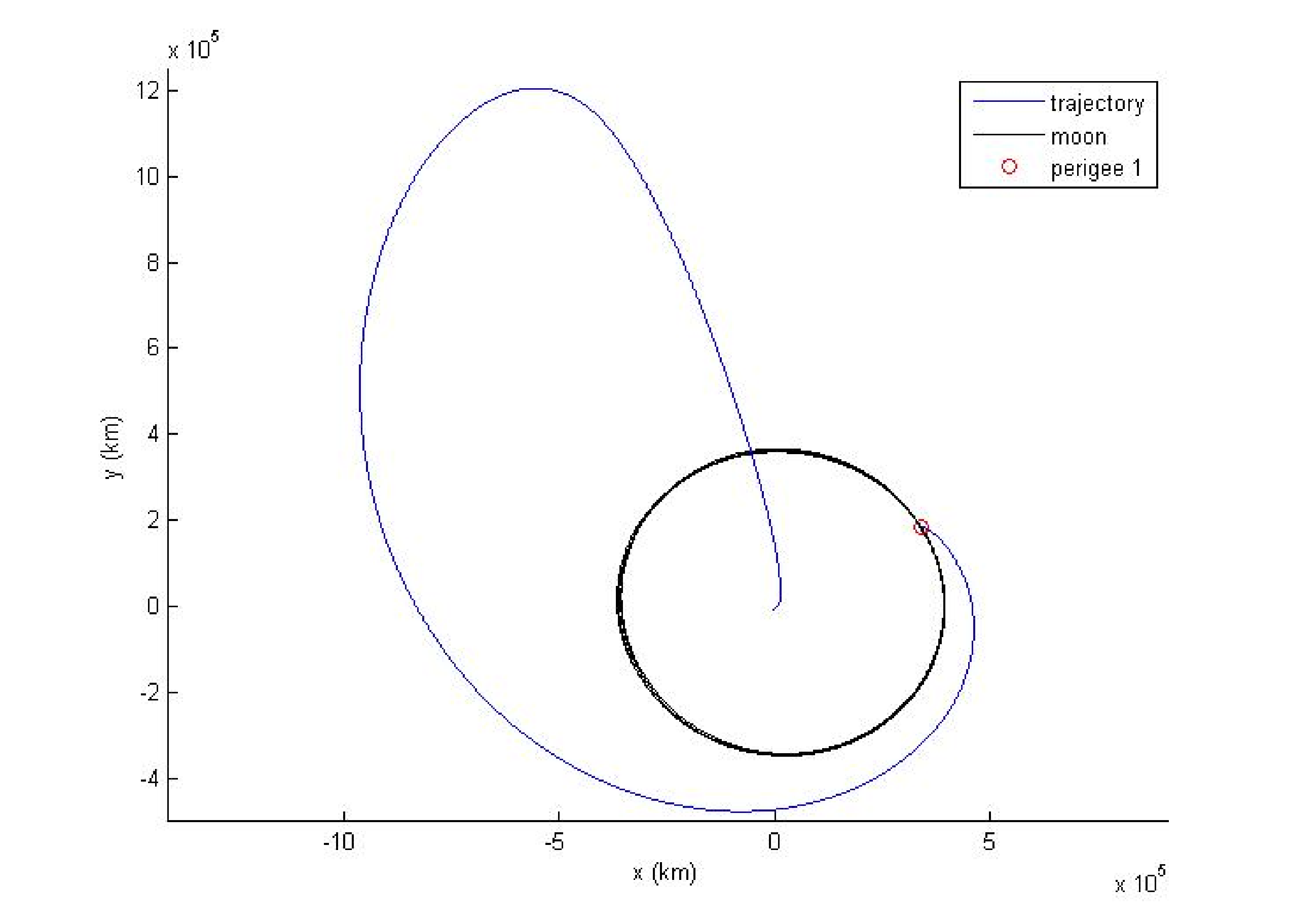}}
\end{center}
\caption{Trajectory from the Earth to the Moon going to the lunar WSB.}
\label{Figure:WSBTransfer}
\end{figure}

\begin{figure}[h!]
\begin{center}
\resizebox{100mm}{!}
{\includegraphics{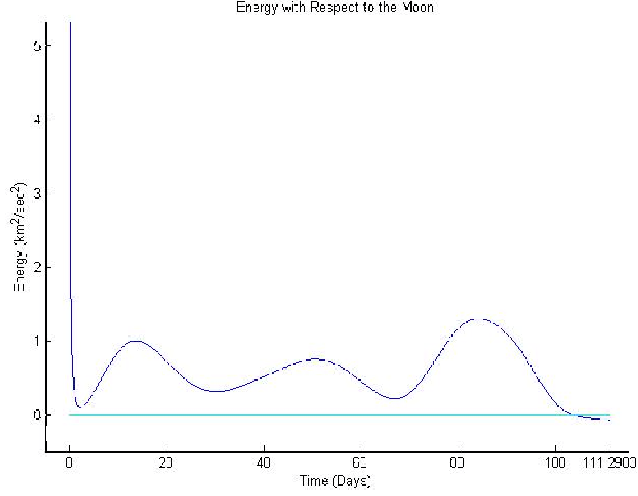}}
\end{center}
\caption{Caption: Variation of the Kepler energy, $K_M$.}
\label{Figure:VinfPlot}
\end{figure}

\item {\it The number of potential targets}. This depends on the probability of finding a star of a given
mass $m_1^*$ in the cluster; we call this $P_{IMF}(m_1^*)$ which describes the 
initial mass function (IMF) of the cluster. Observations of many different star 
clusters (large clusters like the Trapezium, smaller cluster like Taurus 
and even older field stars) find very similar IMFs (down to the Hydrogen burning 
limit at $\sim$ 0.1 $M_{\odot}$; Lada $\&$ Lada 2003 and refereces therein). 
To calculate $P_{IMF}(m_1^*)$ we adopt the IMF of the Trapezium cluster from Lada $\&$ Lada (2003), 
which is characterized by a broken power-law, given by 
$\xi$(M) = $\xi$$_1$M$^{-2.2}$, from 0.6--20 $M_{\odot}$ and 
$\xi$(M) = $\xi$$_2$M$^{-1.1}$, from 0.1--0.6 $M_{\odot}$, 
where $\xi$(M)dM is the number of stars with mass (M, M+dM). [There is a steep decline
into the substellar brown dwarf regime and a possible second peak but we will ignore objects
below the hydrogen burning limit]. To calculate $P_{IMF}(m_1^*)$ 
(square symbols in Figure \ref{Figure:imfcor}), we use a logarithmic binning of masses 
(with d(logM)=0.1), and normalize the distribution to unity, which gives $\xi$$_1$=0.19 
and $\xi$$_2$=0.34. 
\end{enumerate}

\begin{figure}[h!]
\begin{center}
\resizebox{120mm}{!}
{\includegraphics{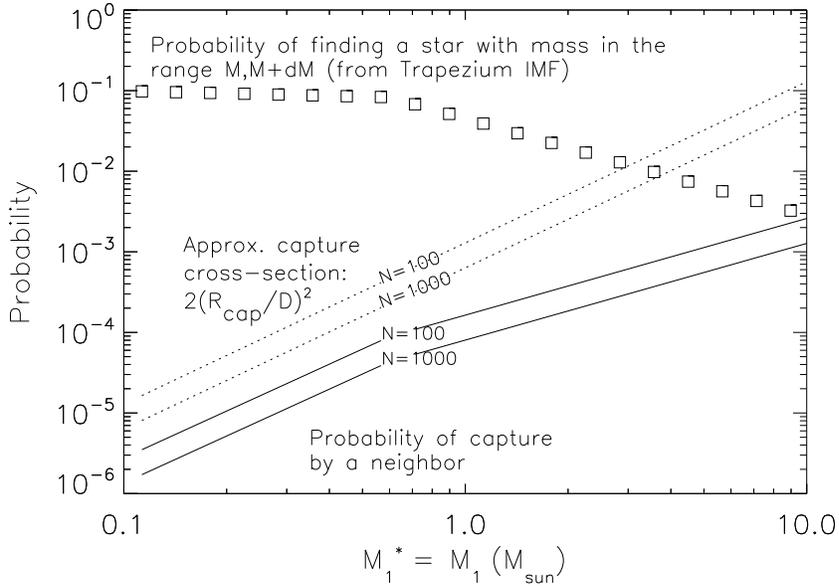}}
\end{center}
\caption{For a given source of remnants, the probability of capture by a neighbor of equal mass, 
shown in $\it{solid}$ line, is given by 
$2(R_{cap}(m_1^*)/D)^2 \times (P_{IMF}(m_1^*))^{2/3}$; where $2(R_{cap}(m_1^*)/D)^2$ is  
shown in $\it{dotted}$ line and $P_{IMF}(m_1^*)$ is the Trapezium IMF normalized to unity, 
shown as $\it{square}$ symbols (Lada $\&$ Lada 2003).}
\label{Figure:imfcor}
\end{figure}

An upper limit to the probability that a remnant escaping from a star of mass $m_1$ (within the range 
described in $\S$5.1) will get captured by a neighboring star of mass $m_1^*$ 
is approximately given by $2(R_{cap}(m_1^*)/D_{m_1^*-m_1})^2$, where $D_{m_1^*-m_1}$ 
is the average distance between two stars of masses $m_1^*$ and $m_1$, respectively.
The simplest case is when both stars have equal masses $m_1^*$ = $m_1$.  
In this case, the average interestellar distance would be $D_{m_1} \sim (1/n_{m_1})^{1/3}$, 
where $n_{m_1}$ the average number density of stars with mass ${m_1}$, 
$n_{m_1} = N_{m_1}/(4/3)\pi R_{cluster}^3$, and $N_{m_1}$ is the total number of stars
in the cluster with mass ${m_1}$, $N_{m_1} = N \times P_{IMF}(m_1)$. 
Because $n_{m_1} = n \times P_{IMF}(m_1)$, 
we get that $D_{m_1} \sim D \times (P_{IMF}(m_1^*))^{-1/3}$. This means that the 
transfer probability between two stars of equal mass is given by 
$2(R_{cap}(m_1^*)/D)^2 \times (P_{IMF}(m_1^*))^{2/3}$, 
where $D$ is the average distance between any two stars in the cluster (regardless
of their mass). 
The resulting capture probability is shown in Figure \ref{Figure:imfcor} as 
a solid line, with values ranging from $10^{-6}$ to $10^{-3}$. The capture 
probability between two planetary systems with solar-type central stars 
(M$_1^*$ = M$_1$ = 1 $M_{\odot}$) is 8.1$\times 10^{-5}$ and 
1.7$\times 10^{-4}$, for a cluster of 1000 and 100 members, respectively. 
\medskip

\begin{center}
5.4 Estimate of the Number of Weak Transfer Events
\end{center}
\medskip

To calculate the total number of remnants that could get transferred between two 
neighboring planetary systems, the capture probability in $\S$5.3 needs to be 
multiplied by the number of remnants, $N_R$, that a planetary system may eject 
before the cluster disperses. 
The main uncertainties in assessing whether weak transfer is a viable method 
for remnant transfer lie in estimating this number. Given the large uncertainties, 
in this sub-section (and the remainder of this paper) we will only estimate $N_R$ for our
Solar System, as it is the only planetary system for which observations and dynamical
models enable us to make an educated estimate. 
\medskip

An estimate of $N_R$ was given in Adams $\&$ Spergel (2005): assuming
that the early solar nebula contained about 50--100 M$_\oplus$ of heavy elements, 
from which about one third was left-over from the planet formation process, and about 
one third of this left-over material was ejected from the planetary system during encounters 
with the giant planets, they estimated that during the time of planet formation, which 
lasted $\sim$ 10 Myr, at least $M_R$ $\sim$ 1 M$_\oplus$ of rocky material could have been 
ejected from a planetary system. Because clusters remain bound during 10--100 Myr, some of this 
rocky material could in principle get captured by other stars in the cluster. Of particular 
interst are the remnants $\sim$10 kg, large enough to shield any potential biological material 
from the hazards of radiation in deep space and from the impact on the surface of 
a terrestrial planet (Horneck 1993; Nicholson et al. 2000; Benardini et al. 2003; 
Melosh 2003).

\medskip

\begin{center}
5.4.1. Estimating $N_R$ from the Oort Cloud
\end{center}
\medskip

An estimate of $N_R$ can be derived from observations and dynamical models of the Oort Cloud comets, 
which are weakly bound to the Solar System and are therefore representative of a population of remnants that may 
have been located on its weak stability boundary, subject to weak escape. 
\medskip

\noindent{\em Oort Cloud formation scenario:}\\
\noindent In a recent paper, Brasser, Duncan \& Levison (2006) proposed the following scenario for Oort Cloud formation. 
Before the gas in the solar protoplanetary nebula dispersed, the gas giant planets Jupiter and Saturn formed, 
and subsequently scattered the Jupiter-Saturn zone planetesimals out to large distances. This process 
happened relatively quickly, before the Sun left its maternal stellar cluster and the cluster gas dispersed. 
At these distances, the planetesimals were subject to the gravitational perturbations of the cluster gas and stars; 
these perturbations cause a Kozai-like effect of coupled eccentricity--inclination oscillations. These long period 
oscillations in a slowly changing gravitational potential of the cluster, resulted in the lifting of the pericenter 
of the planetesimals' orbits beyond the orbit of Saturn. Close encounters with stars, in particular near the center of 
the cluster, also had the effect of increasing the pericenters. 
Planetesimals that achieved pericenters $\gg$ 10 AU were safe from complete ejection from the Solar System, 
and subsequently formed the Oort Cloud.
(This model does not consider the ice giants Uranus and Neptune because they likely formed after 
the solar protoplanetary gas dispersed ($>$10 Myr), and probably after the Sun left the cluster.)  
Brasser et al.~best estimate cluster model that correctly explains the current orbits of Sedna and 2000 CR$_{105}$ and 
that makes Sedna a typical Oort Cloud member consists on a stellar cluster of 284 members, with a total mass in 
stars and gas of 419 M$_{\odot}$, a velocity dispersion of 1.755 km/s, a crossing time of 0.056 Myr, a Plummer 
radius of 0.1 pc and a tidal radius of the Sun of 0.013pc. 
In this model, 2--18\% of the planetesimals in the Jupiter--Saturn region became part of the primordial Oort Cloud. 
\medskip

We adopt the above scenario to calculate $N_R$. First, we estimate the number of planetesimals in 
the Jupiter-Saturn zone, which we assume was the 4--12 AU heliocentric distance zone; we assume that 2--18\% of 
those ended up in the Oort Cloud.
\medskip

\noindent{\em Total mass of solids in the primordial 4-12 AU region:}\\
\noindent To calculate how many planetesimals formed in this region, we estimate the total mass in solids 
(i.e.~excluding H and He) that could have been available for the formation of planetesimals in the 4--12 AU region. 
We adopt the minimum mass solar nebula (MMSN), the minimum mass that the solar protoplanetary disk must have contained in 
order to form all the planets in the Solar System, given by the surface density $\Sigma=\Sigma_0(a/40 AU)^{-3/2}$, 
where the surface density of dust (i.e.~solids) has the coefficient $\Sigma_0=\Sigma_{0d}$ = 0.1 g/cm$^2$, 
(Weidenschilling 1977, Hayashi 1981). Integrating between 4 and 12 AU, we find the total mass in solids, $10^{29}$ g. 
\medskip

\noindent{\em Planetesimal size distribution:}\\
\noindent 
With the above estimate for the total mass in solids, we can calculate the number of planetesimals by adopting a 
planetesimal size distribution function representative of the early Solar System. This size distribution is highly 
uncertain, but can be constrained roughly from observations and coagulation models. 

The current best observational estimate for the size distribution of outer Solar System planetesimals is for the 
trans-Neptunian Kuiper belt bodies studied by Bernstein et al.~(2004).  Theoretical estimates are based on planetesimal 
coagulation models, reviewed in Kenyon et al.~(2007).  Here we summarize these results, and use them to guide our 
estimates for $N_R$.

\begin{itemize}

\item Observations of Kuiper Belt bodies show, broadly, two dynamical classes, the Classical Kuiper Belt (CKB) 
with low inclination, low eccentricity orbits and the Excited Kuiper Belt (EKB) with moderate-to-high orbital 
inclinations and eccentricities.  Bernstein et al.~(2004) find that the size distribution functions of these two 
classes are different at a 96\% confidence level. Each class is well fitted by a `broken' 
power law, $dN/dD \propto$ D$^{-q_1}$ for $D>D_0$ and dN/dD $\propto$ D$^{-q_2}$ if D$<$D$_0$, of 
different power law indices at the small and large sizes: $q_1^{CKB}$$\geq$5.85 and $q_2^{CKB}$$\approx$2.9 for the CKB, 
and $q_1^{EKB}$$\approx$4.3 and $q_2^{EKB}$$\leq$2.8 for the EKB.  
The break in the power laws is at $D_0\approx 100$ km.

\item The CKB has fewer large objects than the EKB; the maximum planetesimal size in the CKB is 60 times smaller 
than the maximum planetesimal size in the EKB.  The EKB contains fewer small objects than the CKB.  
This differences indicate that the Excited population is likely ``older -- has undergone more collisional accretion and 
erosion -- than the Classical population. This suggests that the EKB objects may have formed at smaller heliocentric 
distances than those in the CKB. 

\item For both populations, the size distribution of the small bodies is shallower than expected in a collisional 
cascade, $q_2 < 3.5$, where the latter value is a theoretical estimate for collisional cascades (Dohnanyi 1969).  
The break to a shallower size distribution occurs at $D \leq 100$ km, a size range susceptible to collisional destruction 
(Pan \& Sari 2005).  The observed strong depletion of the small bodies, and the fact that the collisional 
lifetimes in the present-day Kuiper belt are longer than the age of the Solar System, indicates that in the past both the 
CKB and EKB were more massive and richer in small bodies and that the present state is the result of an advanced 
erosional process. 

\item Theoretical models of planetesimal coagulation in the outer Solar System also find a broken power 
law size distribution function, with 
parameter values $q_1\approx 2.7$--$3.3$ and $q_2=3.5$, and a break diameter $D_0\approx1$km (reviewed in 
Kenyon et al.~2007).  These do not match the observations of the present-day Kuiper belt, but may be representative of 
its very early stage during the planet formation era.  

\end{itemize}

Based on the above studies, we adopt three size distribution functions for our calculation of $N_R$.
\begin{itemize}
\item {\it Case A}: $q_1$ = 4.3, $q_2$ = 3.5, $D_0$ = 100 km, 
D$_{max}$ = 2000 km ($\sim$ Pluto's size),
D$_{min}$ = 1 $\mu$m ($\sim$ dust blow-out size).\\
This distribution has the power law index of a collisional cascade at the small size end, and that of the EKB at the 
large size end, with a break diameter consistent with that of the present-day Kuiper belt.  
The rough scenario this case reflects is as follows: In the Jupiter-Saturn zone, the accretion of large 
planetesimals proceeded to make the large-size end similar to that found in the present-day EKB, whereas at the 
small size end, the dynamical stirring by the large bodies produced a classical collisional cascade.

\item {\it Cases B and C} $q_1$ = 3.3 and 2.7, $q_2$ = 3.5, $D_0$ = 2 km,
D$_{max}$ = 2000 km ($\sim$ Pluto's size),
D$_{min}$ = 1 $\mu$m ($\sim$ dust blow-out size):\\   
These size distributions are derived from theoretical coagulation models.

\item {\it Case D}: $q_1$ = 4.3, $q_2$ = 1.1, $D_0$ = 100 km, 
D$_{max}$ = 2000 km ($\sim$ Pluto's size),
D$_{min}$ = 1 $\mu$m ($\sim$ dust blow-out size):\\   
This size distribution represents perhaps the worst case scenario in which the depletion of the small bodies took 
place very early on, before the objects were
thrown into the Oort Cloud.  The parameters here are within the range that Bernstein et al.~(2004) find for the EKB. 
This size distribution is also similar to that found in models for the primordial asteroid belt ($q_1 = 4.5$, $q_2$ = 1.2, 
$D_0$ $\sim$ 100 km, Bottke et al.~2005). 

\end{itemize}

\noindent{\em Number of remnants with masses $>$ 10 kg that could have been subject to weak escape:}\\
With the above estimates of the total mass in solids and of the size distributions, we can finally calculate the 
number of planetesimals with masses $>10$ kg, equivalently, diameter $D > 26$ cm (assuming $\rho$ = 1 g/cm$^{-3}$), 
that would have existed in the 4--12 AU region at the time of Jupiter and Saturn formation. The results are shown 
in Table 1.
The scenario described in Brasser et al. (2006) finds that 2--18\% of these planetesimals would have 
tranferred to the primordial Oort Cloud; these are also the bodies potentially available for weak escape from 
the early Solar System, hence they provide our estimate for $N_R$, also given in Table 1 for each of our cases.  
We find that $N_R$ is in the range $10^{17}$--$10^{20.5}$ for Cases A, B and C (which are based on a collisional 
cascade for the small objects), but is only $10^{7}$--10$^{8}$ for Case D (which has a much shallower size 
distribution for the small bodies, similar to the observed Kuiper belt).  It is clear that $N_R$ is very sensitive 
to the size distribution function, i.e.~power law index used. 
\medskip

\begin{deluxetable}{rcccc}
\tablewidth{0pc}
\tablecaption{Estimated Number of Weak Transfer Events}
\tablehead{
\colhead{Planetesimal Size} & 
\colhead{N$_{D>26cm}$\tablenotemark{b}} &
\colhead{N$_{R}$\tablenotemark{c}} & 
\colhead{N$_{WTE}$\tablenotemark{d}} & 
\colhead{N$_{WTE}$\tablenotemark{d}} \\
\colhead{Distribution\tablenotemark{a}} & 
\colhead{} &
\colhead{} & 
\colhead{(N=100)} & 
\colhead{(N=1000)}}

\startdata
A: $q_1$ = 4.3, $q_2$ = 3.5		& 1.8$\times 10^{21}$	& 3.6$\times 10^{19}$-3.2$\times 10^{20}$	& 6.1$\times 10^{15}$-5.5$\times 10^{16}$	& 2.9$\times 10^{15}$-2.6$\times 10^{16}$\\
$D_0$ = 100 km & & & & \\
B: $q_1$ = 3.3, $q_2$ = 3.5		& 2.7$\times 10^{20}$	& 5.4$\times 10^{18}$-4.9$\times 10^{19}$	& 9.2$\times 10^{14}$-8.3$\times 10^{15}$	& 4.4$\times 10^{14}$-3.9$\times 10^{15}$\\
$D_0$ = 2 km & & & & \\
C: $q_1$ = 2.7, $q_2$ = 3.5		& 8.1$\times 10^{18}$	& 1.6$\times 10^{17}$-1.4$\times 10^{18}$	& 2.7$\times 10^{13}$-2.5$\times 10^{14}$	& 1.3$\times 10^{13}$-1.2$\times 10^{14}$\\
$D_0$ = 2 km & & & & \\
D: $q_1$ = 4.3, $q_2$ = 1.1		& 5.5$\times 10^{8}$		& 1.1$\times 10^{7}$-9.9$\times 10^{7}$		& 1.9$\times 10^{3}$-1.7$\times 10^{4}$		& 8.9$\times 10^{2}$-8.0$\times 10^{3}$\\
$D_0$ = 100 km & & & & \\
\tablenotetext{a}{For all the cases, D$_{max}$ = 2000 km ($\sim$ Pluto's size) and D$_{min}$ = 1 $\mu$m ($\sim$ dust blow-out size).}
\tablenotetext{b}{N$_{D>26cm}$ is the total number of planetesimals with diameter $D > 26$ cm , equivalent to masses $>10$ kg if  $\rho$ = 1 g/cm$^{-3}$.}
\tablenotetext{c}{N$_{R}$ is the expected number of planetesimals with diameter $D > 26$ cm  that populated the primordial Oort Cloud, estimated to be $\sim$2-18\% of  N$_{D>26cm}$.}
\tablenotetext{d}{N$_{WTE}$ is the number of weak transfer events between two solar-mass stars for cluster of N=100 and N=1000 members, calculated as N$_{R}$$\times$Prob$^{cap}$, 
where Prob$^{cap}$ = 8.1$\times 10^{-5}$ for N=1000 and 1.7$\times 10^{-4}$ for N=100. }
\enddata
\end{deluxetable}

\begin{center}
5.4.2 Number of Weak Transfer Events in the Early Solar System
\end{center}
\medskip
Now we estimate the number of weak transfer events from the early Solar System to the nearest solar-type star 
in the cluster (assuming it also harbored a planetary system) by multiplying $N_R$ with the capture probability 
calculated in section 5.3 (see Fig.~5).  For M$_1^*$ = M$_1$ = 1 $M_{\odot}$, the capture 
probability is 8.1$\times 10^{-5}$ for N=1000 and 1.7$\times 10^{-4}$ for N=100. The result is also listed in Table 1.
The number of weak transfer events between two solar-type stars for Cases A, B and C are 
in the range $10^{13.1}$--10$^{16.7}$,
and for Case D is 900--17000. 
As mentioned before, because the location of the renmant inside the 
weak stability boundary does not guarantee capture, these estimates should be regarded as upper limits.
\medskip

\begin{center}
5.5 Implications for Lithopanspermia
\end{center}
\medskip

Because planetesimals could potentially harbor the chemical compounds that constitute the building blocks of life, 
it is of interest that the results above indicate that these materials could have been transferred in significant or 
even large quantities between the Solar System and other solar-type stars in its maternal cluster. For the case of 
binary systems, Adams \& Spergel (2005) found that clusters of  N = 30--1000 could experience billions to trillions of 
capture events among their binary members. However, to consider the transfer 
of microorganisms that developed in the Solar System, a necessary condition would be that life could develop during 
the 10--100 Myr that the cluster remained bounded.  Radiometric measurements of hafnium and tungsten isotopes in 
meteorites indicate that the bulk of the metal-silicate separation in the Solar System occurred within the first 30 
Myr of the solar nebula lifetime, with most of the Earth's core accreted during the first 10 Myr (Yin et al.~2002; 
Kleine et al.~2002). 
It is not known when the conditions for life to develop on Earth were met.  There is evidence, albeit sparse, 
that basic habitable conditions (existence of continents and prevalance of liquid water) were established on 
Earth within $\sim150$ Myr of Solar System formation (e.g., Cates \& Mojzsis 2007).  
The oldest Earth rocks are $\sim3.9$ Gyr old and some authors claim that the first traces of life 
appear just 100 Myr later (Mojzsis et al. 1996), but this result is controversial (Fedo \& Whitehouse 2002).  
Similarly, there is controversy surrounding the finding of modern cyanobacteria in rocks 3.465 Gyr old 
(Schopf 1993; Brasier et al.~2002); undisputed is evidence of cyanobacteria and primitive eucaryotes 
in rocks 2.7 Gyr old (Brocks et al.~1999).  
\medskip

\begin{center}
{\it Transfer Timescales}
\end{center}
\medskip

For the transfer of microorganisms from the Solar System to other systems, we need to consider not only the 
timescale for life to develop,  but also the timescale for weak transfer and how it compares to the timescale for 
microorganism survival in deep space. 

\begin{itemize}
\item {\it Timescale for Ejection}: 
The time, $T_R$,  for a remnant to exit the Solar System, i.e.~to move from its periapsis ($r_p$) with respect to 
the central star to the distance $R_{esc}(m_1)$ can be is estimated using Barker's equation 
which yields the time of flight along a parabolic trajectory,
$$T_R={1\over{2\sqrt{Gm_1}}}( pL + {1\over{3}}L^3 ),$$
where $$L=\sqrt{p} \tan( {\nu \over{2}}), \hspace{.1in} \nu = \arccos
({p\over{R_{esc}}}-1).$$
The variable $\nu$ is the true anomaly, and $p=2r_p(=2\Delta)$ is the semi-latus rectum.  For $p \ll R_{esc}$ and 
using trigonometric identities it can be shown that $L$ varies approximately independently from 
$p: \hspace{.1in} L = \sqrt{2R_{esc}} \sqrt{1- {p \over{2R_{esc}}}} \approx \sqrt{2R_{esc}}$ .
Therefore, the variation of $T_R$ as a function of $r_p$ is negligible, so without loss of generality, we can examine 
the variation of $T_R$ as a function of $R_{esc}$ for a fixed periapsis distance of $r_p \approx 5$ AU: for stellar 
masses of 0.1 $M_{\odot}$, 1 $M_{\odot}$ and 10 $M_{\odot}$, Figure \ref{Figure:escape} shows 
that $R_{esc}$ = 1.8$\times$10$^4$ AU, 1.8$\times$10$^5$ AU and 1.8$\times$10$^6$ AU, respectively, 
which yields $T_R \approx$ 0.6, 6 and 60 Myr, respectively. 

\item {\it Timescale for interstellar transfer}:
For a remnant moving at the low velocity of 0.1 km/s, as required by the weak transfer mechanism, 
it will take about 3.3--4.8 Myr to reach a neighboring star located at $D \approx 7\times$10$^4$--$10^5$ AU. 

\item {\it Timescale to land on a terrestrial planet}:
Because it is not required that the captured remnant approaches the star in the ecliptic plane, multiple 
periapsis passages about the target star would be needed before the remnant can collide with a planet. This can 
take of the order of tens of millions of years. 

\item {\it Timescale for erosion from interplanetary dust}: 
In the interplanetary dust environment found in the Solar System, a meter-size rock will be ablated and eroded 
on a timescale of about 0.02--0.23 Myr, short compared to the time it takes to be ejected from the 
system (about 4 Myr; Napier 2004). It is likely that the early Solar System was significantly more dusty than it 
is today, thus survival times of remnants would have been even shorter, and
much less than the several Myr transfer timescales.
\end{itemize}

\medskip

Even if we were to assume that life arose on Earth before the stellar cluster dispersed, given the ejection 
and transfer timescales described above and the survival timescale of dormant microoganisms, it seems unlikely 
that microorganisms could have been transferred even among the closest neighbors in the cluster via the weak transfer 
mechanism described here. 
However, our results that significant quantities of solid material could have been transferred via weak transfer 
between nearest neighbors in the Sun's birth cluster, are worth further investigation for the exchange of chemical 
compounds that constitute the building blocks of life and which could potentially have significantly longer survival 
lifetimes. 

\medskip

\begin{center}
6. CONCLUSIONS
\end{center}
\medskip
\noindent

Could life on Earth have been transferred to other planetary systems within the first few Myr of the Solar System 
evolution, when the Sun was still embedded in its maternal aggregate?; and vice versa, could life on Earth have been 
originated beyond the boundaries of our Solar System?
\medskip

In this paper we have described a dynamical mechanism that yields very low velocity chaotic escape of remnants from 
a planetary system using parabolic trajectories, and its reverse process of chaotic capture. These two processes provide 
a mechanism for minimal energy transfer of remnants between planetary systems.  We have applied this mechanism to the 
problem of planetesimal transfer between planetary systems in an open star cluster, where the relative velocities 
between the stars are sufficiently low ($\sim$ 1 km/s) to allow slow escape and capture. Based on geometrical 
considerations, we have estimated upper limits to the probability of transfer that depend on the cluster properties 
and the masses of the source and target stars. Using these probabilities, adopting the Oort Cloud formation models 
from Brasser et al.~(2006), and adopting a range of planetesimal size distributions derived from observations and 
theoretical models (Bernstein et al.~2004 and Kenyon et al.~2007, respectively), we estimate the number of 
weak transfer events from the early Solar System to the nearest solar-type star in the cluster (assuming it also 
harbored a planetary system). This estimate is most sensitive to the power law index of the size distribution of small 
bodies: the number of weak transfer events could be could be as large as $10^{13.1}$--$10^{16.7}$ if the size 
distribution of the small bodies follow a classical collisional cascade, or as small as 900--17000 if we adopt a 
shallower power law index for the size distribution of the small bodies.  
To determine whether the weak transfer process described in this paper could have been a viable mechanism for the 
transfer of remnants between the Solar System and other stars in the cluster, further progress needs to be made in 
the understanding of the dynamical and collisional history of the early Solar System. 
  
\medskip
 
Comets and asteroids have been suggested to be a possible source of the chemical compounds that constitute the 
building blocks of life on Earth.  The results summarized above are therefore of interest because they indicate 
that there is the possibility (depending on the shape of the size distribution) that planetesimal material could have 
been transferred via the weak transfer mechanism in large quantities between the Solar System and other stars in its 
maternal cluster.

\medskip\medskip\medskip

\noindent
{\it Acknowledgments}\\

E.B. acknowledges support from the NASA SMD/AISR program. 
A.M.M. is under contract with the Jet Propulsion Laboratory (JPL) funded by NASA through the Michelson 
Fellowship Program. JPL is managed for NASA by the California Institute of Technology. A.M.M. is also supported by the 
Lyman Spitzer Fellowship at Princeton University.
R.M.~acknowledges support from NASA's Outer Planets Program and from the Life and Planets Astrobiology 
Center (LaPLaCe) of the University of Arizona which is supported by the NASA Astrobiology Institute.

\end{document}